\documentstyle[twocolumn,aps,epsfig]{revtex}

\newcommand{\bqa}{\begin{eqnarray}}
\newcommand{\eqa}{\end{eqnarray}}
\newcommand{\nn}{\nonumber \\}

\begin{document}
\draft 
\twocolumn[\hsize\textwidth\columnwidth\hsize\csname @twocolumnfalse\endcsname
\title{Scaling behavior in the optical conductivity of 
the two dimensional systems of strongly correlated electrons
based on the U(1) slave-boson approach to the t-J Hamiltonian}

\author{Jae-Hyeon Eom, Sung-Sik Lee, Ki-Seok Kim and Sung-Ho Suck Salk$^a$}
\address{Department of Physics, Pohang University of Science and Technology,\\
Pohang, Kyoungbuk, Korea 790-784\\
$^a$ Korea Institute of Advanced Studies, Seoul 130-012, Korea\\}
\date{\today}

\maketitle

\begin{abstract}
The U(1) holon-pair boson theory of Lee and Salk(Phys. Rev. B {\bf 64}, 052501 (2001)) is applied to investigate the quantum 
 scaling behavior of optical conductivity in the two dimensional systems of strongly correlated electrons.
We examine the role of both the gauge field fluctuations and spin pair excitations on the $\omega / T $ 
scaling behavior  of the optical conductivity.
It is shown that the gauge field fluctuations but not the spin pair excitations are responsible
for the scaling behavior in the low frequency region $\omega/T <<1$. 
Importance for the contribution of the nodal spinons to the Drude peak is discussed.
It is shown that the $\omega / T$ scaling behavior is manifest in the low frequency region  at low hole 
concentrations  close to a critical concentration at which superconductivity arises at $T=0K$.
\end{abstract}
\vskip2pc]
\narrowtext
\newpage

\section{INTRODUCTION}

A quantum phase transition(QPT) is a zero temperature(T=0) phase transition induced by quantum  fluctuations. 
 The strength of the quantum fluctuations is
controlled by `external' parameters such as magnetic field, pressure, phase stiffness, doping and etc.
 These external parameters correspond to some coupling constant $g$ in the Hamiltonian and the QPT occurs 
as a result of the change of the ground state at a critical value $g_c$.
The QPT can be a generic feature of the strongly correlated systems; superconductor-insulator transitions,
metal-insulator transitions, integer and fractional quantum hall transitions, magnetic transitions\cite{SACHDEV}. 
Various studies concerned with QPT in the high $T_C$ cuprates have been made:
electron fractionalization\cite{SENTHIL_FRAC,ICHINOSE,RIBEIRO,LEE_NAGAOSA,ANDERSON_SCS},
magnetic properties\cite{PRELOVSEK,SACHDEVMAGNETISM,KEIMER,SOKOL,AEPPLI},
transport properties\cite{SACHDEVCON,MAREL_TRANSP,DEVEREAUX},
time-reversal symmetry breaking\cite{KAMINSKI,SANGIOVANNI,CHAKRAVARTY_TSB,VOJTA_TSB}
and phenomenological QCP near the optimal doping concentration\cite{VARMA,LORAM,VALLA,SACHDEVPRB}.
One of the most interesting studies in the QPT is  to see if there exists universal $\omega / T$ scaling behavior of response 
functions in the quantum critical region of $ |g-g_c|/T << 1 $\cite{SACHDEV,TSVELIK}.
In the quantum critical region the energy gap $\Delta$ satisfies $\Delta \sim T$ and the response function 
shows the $\omega / T$ scaling behavior; for example $\sigma (\omega, T, g) \sim T^{d-2} \Sigma (\omega /T , \Delta / T)
\sim  T^{d-2} \Sigma (\omega /T )$ with $\sigma$, the optical conductivity at frequency $\omega$\cite{SACHDEV,SACHDEVCON}.
In the high $T_C$ cuprates, the external parameter can be the hole doping concentration $x$.
With the increase of the hole doping concentration at zero temperature, the high $T_C$ cuprates show diversified phases of 
antiferromagnetism, pseudogap and superconductivity.
Hole doping to the parent compound of the antiferromagnetic(AF) Mott insulator induces frustration
of the AF order and as a result the pseudogap(PG) phase occurs.
Further hole doping results in the superconductivity(SC) and the superconducting transition temperature $T_C$ increases
to a maximum at an optimal concentration, beyond which $T_C$ decreases until SC disappears,
thus showing an arch shape of the superconducting transition temperature over a range of doping.
In this paper we investigate the $\omega / T$ scaling behavior of the in-plane optical conductivity based on the U(1) 
slave-boson theory of Lee and Salk\cite{SSLEE} which has been successful in reproducing various observations, 
 including the arch-shaped superconducting transition temperature in the phase diagram of high $T_C$ cuprates. 
This theory is different from other previous slave-boson theories\cite{KOTLIAR,FUKUYAMA,WEN,GIMM} in that the
Heisenberg interaction term in the slave-boson representation contains the importance of coupling between the 
charge and spin degrees of freedom. 
Both the U(1) and SU(2) theories of Lee and Salk\cite{SSLEE} showed that the hump structure in the optical conductivity
is originated from antiferromagnetic spin fluctuations of short range, including the spin singlet pair excitations\cite{LEKS,ELKS}.
In these studies we found that the gauge field fluctuations affect the Drude peak but not the hump structure.
In the present work we show that the optical conductivity reveals the $\omega/T$ scaling behavior 
under the influence of  the gauge field fluctuations which enforce the back-flow condition.
The importance of the nodal quasi-particle excitations in the cold spot of the Brillouin zone 
for the quantum phase transition phenomena is discussed.

\section{THEORY}

\subsection{DERIVATION OF THE OPTICAL CONDUCTIVITY IN THE U(1) SLAVE-BOSON THEORY}

To reveal differences with other proposed slave-boson theories we briefly present only the  
rudimentary part of the U(1) slave-boson theory proposed by Lee and Salk\cite{SSLEE} 
and the derivation of the optical conductivity in the U(1) slave-boson theory\cite{LEKS,ELKS}.
The t-J Hamiltonian in the presence of the external 
electromagnetic field ${\bf A}$
is written,
\bqa
H &=& -t \sum_{<i,j>} ( e^{i A_{ij}} \tilde{c}^{\dagger}_{i \sigma} 
\tilde{c}_{j \sigma} + H.C ) \nn
&&+J \sum_{<i,j>}( {\bf S}_i \cdot {\bf S}_j - 
\frac{1}{4} {n}_{i} {n}_{j}) 
- \mu \sum_{i, \sigma}c^\dagger_{i \sigma} c_{i \sigma},
\label{t-J_Hamiltonian}
\eqa 
with ${\bf S}_i = 
\frac{1}{2} \sum_{\alpha \beta }c_{i\alpha}^{\dagger} {\sigma}_{\alpha \beta} c_{i\beta} $.
Here $ A_{ij}$ is the external electromagnetic field,
$\tilde{c}_{i \sigma}( \tilde{c}^{\dagger}_{i \sigma})$, the electron annihilation(creation) operator  at each site,
$ {\sigma}_{\alpha \beta} $, the Pauli spin matrix and $\mu$, the chemical potential.
It is noted that the t-J Hamiltonian is the effective Hamiltonian in the large U(on-site Coulomb repulsion energy) limit 
of the Hubbard model.
Thus the electron cannot hop to an occupied site.  
Rewriting the electron operator as a composite of spinon($f$) and holon($b$) operators, 
\bqa
c_{i \sigma} = f_{i \sigma} b^{\dagger}_i,
\label{qcp_c_slaveboson_ref}
\eqa
the partition function is written as
\bqa
Z = \int 
{\cal D}f 
{\cal D}b
{\cal D}\lambda
 e^{-\int_{0}^{\beta} d \tau {\cal L}},
\eqa
where ${\cal L}= \sum_{i}\left( \sum_{\sigma}f^{*}_{i \sigma} \partial_{\tau} f_{i \sigma} 
+  b^{*}_{i} \partial_{\tau} b_{i}   \right)+ H_{t-J}$ is the Lagrangian
with $H_{t-J}$, the U(1) slave-boson representation of the above t-J Hamiltonian(Eq.(\ref{t-J_Hamiltonian})), 
\bqa
H_{t-J} & = & -t\sum_{<i,j>,\sigma}(e^{i A_{ij}}
f_{i\sigma}^{\dagger}f_{j\sigma}b_{j}^{\dagger}b_{i} 
+ c.c.) \nonumber \\
&& -\frac{J}{2} \sum_{<i,j>} b_i b_j b_j^{\dagger}b_i^{\dagger}
(f_{i\downarrow}^{\dagger}f_{j\uparrow}^{\dagger}-f_{i\uparrow}^ {\dagger}
f_{j\downarrow}^{\dagger})(f_{j\uparrow}f_{i\downarrow}-f_{j\downarrow}
f_{i\uparrow}) \nn
&& - \mu\sum_{i,\sigma}  f_{i\sigma}^{\dagger}f_{i\sigma}
+ i\sum_{i} \lambda_{i}(\sum_{\sigma}f_{i\sigma}^{\dagger}f_{i\sigma}+b_{i}^{\dagger}b_{i} -1).
\label{Hamiltonian_fb}
\eqa
Here $\lambda_i$ is the Lagrange multiplier field which enforces the single occupancy constraint.

Applying the Hubbard-Stratonovich transformations involving hopping, spinon pairing and holon pairing
orders we obtain  the partition function, 
\bqa
Z = \int
{\cal D}f 
{\cal D}b
{\cal D}\chi
{\cal D}\Delta^f
{\cal D}\Delta^b
{\cal D}\lambda
e^{-\int_{0}^{\beta} d \tau {\cal L}_{eff}},
\label{free_energy_final}
\eqa
where ${\cal L}_{eff} = {\cal L}_0 + {\cal L}_f + {\cal L}_b $ is
the effective Lagrangian with
\small{
\bqa
{\cal L}_0 =&& \frac{J(1- x)^2 }{2} \sum_{<i,j>} \Big\{ 
|\Delta^f_{ij}|^2 + 
\frac{1}{2} |\chi_{ij}|^2 + \frac{1}{4} \Big\} \nn
&&+ \frac{J}{2} \sum_{<i,j>}
|\Delta^f_{ij}|^2 (|\Delta^b_{ij}|^2 + x^2), 
\label{qcp_orderparameter_u1}
\eqa } \normalsize
 for the order parameter Lagrangian and
\small{
\bqa
{\cal L}_f  
=&& \sum_{i,\sigma} f^\dagger_{i \sigma}( \partial_{\tau}- \mu^f ) f_{i \sigma} \nn
&& -\frac{J(1- x)^2}{4}\sum_{<i,j>,\sigma}
\Big\{ \chi^*_{ij} f^\dagger_{i \sigma}f_{j \sigma} + c.c. \Big\} \nn
&&- \frac{J(1- x)^2}{2} \sum_{<i,j>}
\Big\{ {\Delta^f}^*_{ij} ( f_{i  \downarrow} f_{j  \uparrow}
- f_{i  \uparrow} f_{j  \downarrow}) 
+ c.c. \Big\}
\label{qcp_spinon_u1}
\eqa   } \normalsize
for the spinon sector and  
\small{
\bqa
 {\cal L}_b =&& 
\sum_{i} b^\dagger_{i} (\partial_{\tau} - \mu^b) b_{i}
 -t \sum_{<i,j>}
\Big\{
\chi^*_{ij} b^\dagger_{i} b_{j} + c.c.
\Big\} \nn
&&-\frac{J}{2} \sum_{<i,j>}
|\Delta^f_{ij}|^2 \Big\{
{\Delta^b_{ij}}^* b_i b_j + c.c.
\Big\}
\label{qcp_holon_u1}
\eqa }\normalsize
for  the holon sector.
Here  $\chi$, ${ \Delta }^f$ and ${ \Delta }^b$ are
the hopping, spinon pairing and holon pairing order parameters
respectively. $\mu^f (\mu^b)$ is the spinon(holon) chemical potential and $x$, the hole concentration.

We compute the optical conductivity $\sigma (\omega)$ directly from the current response function $\Pi (\omega)$,
\small{
\bqa
\sigma ( \omega ) 
= \left. \frac{ \partial J_x (\omega) }{ \partial E_x (\omega)} \right|_{ E_x =0} 
=- \frac{1}{ i \omega } 
\left. \frac{ \partial^2 F  }{ \partial {A_x}^2} 
\right|_{A_x =0} 
= \frac{ \Pi_{xx} (\omega) }{ i \omega},
\eqa}\normalsize
where the induced current $J_x$ is chosen in the x direction by considering the case of isotropic current,
and $ E_x (\omega)$, the external electric field
with frequency $\omega$. Here $F = - k_B T \ln Z$ is the free energy and $A_x (\omega)$, the external electromagnetic field.
In the slave-boson theory the current response function $\Pi (\omega)$ is given solely by
the holon current response function because the spinon has no electric charge. 
In this case the U(1) gauge symmetry is broken and the current is not conserved. 
The hopping order parameter $\chi_{ij} = |\chi_{ij}|e^{a_{ij}}$ defines the gauge field,
 $a_{ij}= \partial_{ij}\theta = \theta_i - \theta_j$ and the gauge field fluctuations leads to the
back flow condition for the spinon and holon currents $J^f + J^b = 0$. 
Allowing the interplay between the charge and spin degrees of freedom originated from the 
 kinetic energy term of the t-J Hamiltonian,
Ioffe-Larkin composition rule for the electron current response function\cite{IOFFE} is obtained,
\bqa
\Pi &=& 
         \frac{\Pi^f \Pi^b}{\Pi^f + \Pi^b}, 
\label{qcp_ioffelarkin_rule}
\eqa
where  $\Pi^{f}$ and $\Pi^{b}$ are the spinon and holon current response functions respectively.
The effects of  spin degrees of freedom are manifested  through the antiferromagnetic spin fluctuations 
which appear in the Heisenberg exchange coupling term. 
This exchange interaction leads to the antiferromagnetic spin fluctuations of short range order(including the spin singlet pair).
Thus to incorporate the spin fluctuations into the current response function we  
allow amplitude fluctuations of the spinon singlet pairing order parameter $|{ \Delta }^f|$.
Then we obtain the current response function\cite{LEKS,ELKS}, 
\bqa
\Pi &=& 
	 \frac{\Pi^f \Pi^b}{\Pi^f + \Pi^b} 
+ \frac{ 
\left( \Pi^b_{a \Delta} - 
       \frac{\Pi^b_{a \Delta} + \Pi^f_{a \Delta}}
       {\Pi^b + \Pi^f} \Pi^b    
\right)^2 }
{   2  \frac{   ( \Pi^b_{a \Delta } + \Pi^f_{a \Delta } )^2}
      { \Pi^b + \Pi^f }  
- ( \Pi^0_{\Delta \Delta} + \Pi^b_{\Delta \Delta} + \Pi^f_{\Delta \Delta} )   }, 
\label{Pi}
\eqa
where  $ \Pi^f_{a \Delta} = - \frac{ \partial^2 F^{f} }{ \partial a \partial |\Delta^f| }$
and $ \Pi^b_{a \Delta} = - \frac{ \partial^2 F^{b} }{ \partial a \partial |\Delta^f| }$ 
are the spinon(holon) response function associated with both the gauge fields and the spinon pairing field
and $\Pi^f_{\Delta \Delta}$, $\Pi^b_{\Delta \Delta}$ and  $\Pi^0_{\Delta \Delta}$, 
the response function associated with the spinon pairing field.
It is reminded that the first term of Eq.(\ref{Pi}) represents the Ioffe-Larkin rule 
contributed only from  the kinetic energy term involved with gauge field fluctuations, and the second term,  from the spin fluctuations.
The second term comes from the coupling between the charge and spin degrees of freedom in which the antiferromagnetic 
spin fluctuations are embedded.
Earlier we have shown that the first term solely contribute to the Drude peak and the second term,  
to the hump structure of the optical conductivity\cite{LEKS,ELKS}.

\subsection{SCALING BEHAVIOR OF THE CONDUCTIVITY IN THE TWO DIMENSION}

The dimensional analysis shows that the conductivity $\sigma (\omega)$ has scaling dimension of $d-2$ and
the conductivity  has the scaling form\cite{SACHDEV,SACHDEVCON},
\bqa
\sigma ( \omega , T, g ) = \frac{Q^2}{\hbar}\left(\frac{k_B T}{\hbar c} \right)^{d-2}
                    \Sigma \left(\frac{\hbar \omega }{ k_BT} , \frac{\Delta}{k_B T } \right),
\label{qcp_scaling_conductivity}
\eqa 
where $ \Sigma ( \hbar \omega/ k_B T , \Delta / k_B T  )$ is the scaling function, with $\Delta \sim \xi^{-z}$, the energy gap
and $Q$, the charge of the charge carrier. 
Here $\xi$ is the correlation length and $z$, the dynamics critical exponent.
In the critical region of $\Delta \sim k_B T$, the conductivity shows universal scaling behavior,
\bqa
\sigma ( \omega ) = \frac{Q^2}{\hbar}\left(\frac{k_B T}{\hbar c} \right)^{d-2}
                    \Sigma \left(\frac{ \hbar \omega }{k_B  T}  \right).
\label{qcp_scaling_conductivity_2}
\eqa 
In this work we are concerned with the two dimensional system of the high $T_C$ cuprates 
and the conductivity has no scaling dimension in temperature.
Thus we obtain
\bqa
\sigma(\omega) \sim \Sigma (\frac{\omega}{T})
\label{qcp_scaling_conductivity_final}.
\eqa
Numerical calculations of the scaling function by Damle and Sachdev showed a bell-shaped feature 
in the plot of $\Sigma_I'$ vs. $\omega / T$, where $\Sigma_I'$ is the real part of the scaling function 
for low frequency $\omega /T<< 1$ in the two spatial dimension\cite{SACHDEVCON}.

\section{COMPUTED RESULTS OF IN-PLANE OPTICAL CONDUCTIVITY}

\begin{figure}[h]
\centerline{\epsfig{file=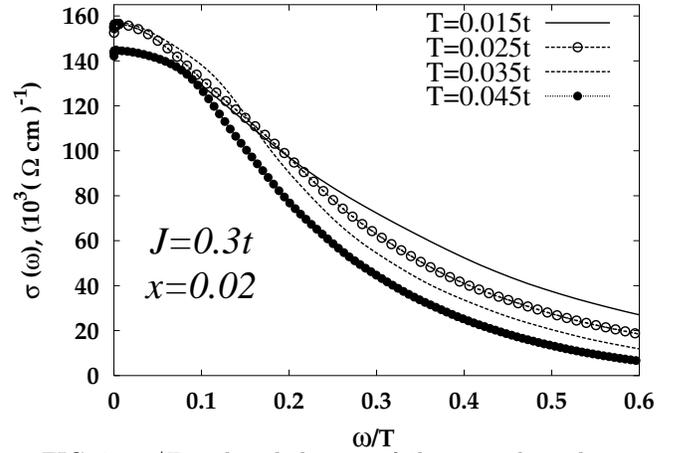, width=9cm,angle=0}}
\caption{$\omega / T$ scaling behavior of the optical conductivity with the inclusion of both the gauge field fluctuations and
the spin pair excitations based on the U(1) slave-boson theory at $x=0.02$}
\label{qcp_omega_T_u1_gauge_Delta_f}
\end{figure}

Here we present the predicted $\omega / T$ scaling behavior of the conductivity $\sigma ( \omega , T )$ 
with the inclusion of both the gauge field fluctuations and the spin pair excitations based on the U(1) slave-boson theory.
In order to see whether the quantum phase transition occurs at low hole doping and low temperature we choose
$x=0.02$ near a critical doping at which the superconducting transition begins to occur at $T=0K$\cite{SSLEE}. 
The temperature range is chosen between 0.015t to 0.045t which is above $T_C$(0.013t) and below $T^*$(0.07t), 
based on the predicted phase diagram obtained by Lee and Salk\cite{SSLEE}.
Fig.\ref{qcp_omega_T_u1_gauge_Delta_f} shows the $\omega / T$ scaling behavior of the optical conductivity
for temperatures between 0.015t and 0.045t.
 We see that the $\omega / T$ scaling behavior becomes markedly clear for  $\omega /T << 1$
in the region of low temperature as shown in Fig.\ref{qcp_omega_T_u1_gauge_Delta_f}.
Encouragingly we obtained a bell shape in the plot of $\sigma(\omega)$ vs. $\omega/T$ in agreement with 
the bell shape feature obtained by Damle and Sachdev(see Fig.6 in Ref.\cite{SACHDEVCON}).

\begin{figure}[h]
\centerline{\epsfig{file=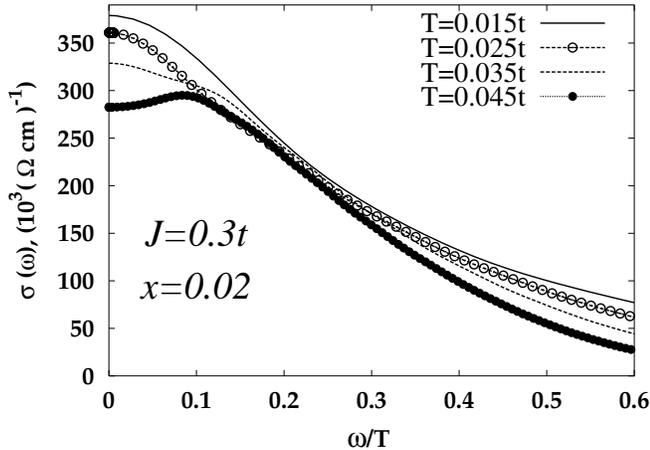, width=9cm,angle=0}}
\caption{Deviation from the $\omega / T$ scaling behavior of the optical conductivity in the low frequency region
with the exclusion of  gauge field fluctuations}
\label{qcp_omega_T_u1_no_fluc}
\end{figure}

Although not shown here we found that the neglect of the spin singlet pair excitations(spinon singlet pair order) 
does not affect the scaling behavior of the conductivity.
Realizing that such excitations affect only the hump structure, but not the Drude peak in the optical conductivity, 
we note that the spin fluctuations contribute only to higher energy(frequency) charge dynamics involved with
a mid-infrared band in the high $T_C$ cuprates. 
For further analysis on the scaling behavior of the optical conductivity we show the $\sigma(\omega)$ vs. $\omega / T$
with the exclusion of gauge field fluctuations, that is, the conductivity contributed 
only from the bare kinetic energy term of the holons(Fig.\ref{qcp_omega_T_u1_no_fluc}). 
Comparison of  Fig.\ref{qcp_omega_T_u1_gauge_Delta_f} and Fig.\ref{qcp_omega_T_u1_no_fluc} shows that the 
scaling behavior is originated from the kinetic energy term involved with the gauge fluctuations which satisfy the back-flow condition.
As shown in Fig.\ref{qcp_omega_T_u1_no_fluc} deviations from the universal scaling behavior become increasingly larger
as $\omega / T $ decreases to 0. 
We find that the nodal quasi-particle excitations for the Drude peak 
in the cold spot of the Brillouin zone are important for the $\omega / T$ scaling behavior of a bell shape, as 
shown in Fig.\ref{qcp_omega_T_u1_gauge_Delta_f}.

Both the spin and charge degrees of freedom are revealed through the back-flow condition resulting from the gauge field fluctuations,
$J^f+J^b=0$ with $J^f$, the spinon current and $J^b$, the holon current.
In our previous study of the optical conductivity\cite{LEKS,ELKS}, the mid infra-red hump in the optical conductivity 
was found to result from the antiferromagnetic spin fluctuations of short range, including the spin singlet
pair excitations, i.e., the shortest possible antiferromagnetic correlation length 
while the Drude peak is from the nodal point(the cold spot) in the Brillouin zone. 
Therefore we conclude that only the nodal quasi-particle excitations in the cold spot contribute to the scaling behavior of the conductivity 
at low frequencies and play a dominant role in the quantum critical phenomena.
As shown in Fig.\ref{qcp_omega_T_u1_gauge_Delta_f}, the $\omega / T$ scaling behavior survives in a relatively narrow 
range of temperature below the pseudogap temperature $T^*$.

\section{SUMMARY}

In the present study, we investigated the $\omega / T$ scaling behavior in the conductivity 
of the two dimensional systems of strongly correlated electrons based on the holon-pair boson theory
of Lee and Salk\cite{SSLEE}.  
The universal scaling behavior is predicted in a small range of temperature above $T_C$ but
 below $T^*$ as long as hole concentration is limited to a substantially small value in the underdoped region.
By observing the disappearance of the scaling behavior in the low frequency region with the exclusion of
the gauge field fluctuations(which enforce the back-flow condition),
we conclude that  the nodal quasi-particle excitations in the cold spot of the Brillouin zone contribute to 
the $\omega / T$ scaling behavior in the conductivity.
The hot spot quasi-particles are seen to be irrelevant to the quantum (phase transition) critical behavior at low temperature.
We find that the spin singlet pair excitations do not contribute to the $\omega / T$ scaling behavior at $\omega << T$.

\end{document}